\documentclass[aps,prd,reprint,superscriptaddress,nofootinbib,nobibnotes]{revtex4-2}

\usepackage{amsmath,amssymb,bm}
\usepackage{graphicx}
\usepackage{siunitx}
\usepackage{hyperref}
\usepackage{comment}

\begin{document}

\title{Higgsino dark matter compatible with the LUX-ZEPLIN high-energy\\ nuclear-recoil event and IceCube constraints}

\author{Katherine Freese}
\email{ktfreese@utexas.edu}
\affiliation{Weinberg Institute for Theoretical Physics, Department of Physics, The University of Texas at Austin, Austin, TX 78712, USA}
\affiliation{The Oskar Klein Centre, Department of Physics, Stockholm University, AlbaNova, SE-106 91 Stockholm, Sweden}
\affiliation{Nordita, Stockholm University and KTH Royal Institute of Technology, Hannes Alfvéns väg 12, SE-106 91 Stockholm, Sweden}
\author{Dionysios P. Theodosopoulos}
\email{d.theodosopoulos@utexas.edu}
\affiliation{Weinberg Institute for Theoretical Physics, Department of Physics, The University of Texas at Austin, Austin, TX 78712, USA}

\date{\today}

\begin{abstract}
We examine Higgsino dark matter (DM) as a potential explanation of the $248~\mathrm{keV}$ nuclear-recoil event of interest reported by the LUX-ZEPLIN (LZ) experiment,
taking into account IceCube constraints on DM annihilation in the Sun.
In the nearly pure Higgsino limit, two neutral Majorana states separated by a small mass splitting $\delta$ are connected by an off-diagonal coupling to the $Z$ boson, providing a natural realization of inelastic DM.
Using a likelihood analysis, we identify 90\% confidence regions compatible with the LZ event and consistent with the null results of IceCube searches, both with and without 
including a constraint from the absence of events in the LZ high-energy sideband.
In our LZ analysis, we consider  the DM distribution in
the Standard Halo Model alone as well as possible enhancements of the high-speed tail induced by the Large Magellanic Cloud (LMC).
We present four variants of bounds from IceCube on Higgsino dark matter, depending on various assumptions on cooling channels in the Sun,
motivated by severe theoretical uncertainties about Higgsino-nucleon elastic scattering (unimportant for direct detection but crucial for IceCube bounds).
Whereas the thermal Higgsino with a mass of $1.1~\mathrm{TeV}$ remains only marginally compatible with both the LZ event and the IceCube limits when the LMC-induced high-speed tail is included and both elastic cooling in the Sun and the LZ high-energy sideband constraint are neglected, higher mass Higgsinos are excellent candidates for the LZ event in light 
of IceCube constraints.
Depending on the halo distribution, cooling interactions, and whether the LZ high-energy sideband is included, the compatible regions in the Higgsino parameter space $(m_\chi,\delta)$ extend from masses of $\sim2~\mathrm{TeV}$ to hundreds of TeV or higher, with mass splittings ranging approximately from $350$ to $525~\mathrm{keV}$.

\end{abstract}

\maketitle

\section{Introduction}

Although Dark Matter (DM) constitutes most ($\sim 85\%$) of the mass of the Universe, its nature remains a mystery.
One of the best-motivated candidates is Weakly Interacting Massive Particles (WIMPs), which are proposed to exist within several scenarios from Beyond the Standard Model (BSM) of particle physics.   


Direct detection laboratory experiments, originally proposed in the 1980s 
 \cite{Goodman:1984dc,Drukier:1986tm},
 are designed to probe the interactions of WIMP dark matter with atomic nuclei 
in terrestrial detectors situated  deep underground.
Despite decades of experimental searches with continuously improving sensitivity to rare dark matter interactions,
the particle nature of dark matter remains unsolved.
Over the years there have been hints of detection, but conclusive discovery remains elusive.
The DAMA/LIBRA experiment observed an annual modulation of the signal in NaI detectors
consistent with a WIMP interpretation\cite{Bernabei:2013xsa}, but these results are in tension with recent experiments made of the same  material \cite{ANAIS-112:2025fne}.

Recently, a single unexplained event found by the LUX-ZEPLIN (LZ) Collaboration has generated a great deal of excitement, 
as potentially due to a dark matter particle interacting with the detector.
The LZ experiment, consisting of xenon nuclei with a $2.84$ tonne-year exposure, extended dark matter searches to a regime not previously studied,
namely searches with this high exposure out to a higher nuclear-recoil energy range, with sensitivity to $5.4$--$270~\mathrm{keV}$ \cite{LZ:2026axp} recoils.  
They found an event 
of interest  with a nuclear recoil of 248 $\pm$ 23 (stat) $\pm$ 23 (sys) keV.  Rare background processes and detector effects  were studied in detail
and shown not to be consistent with this the event. 
To quantify its significance, they used
 a profile likelihood ratio test, and found the background only hypothesis to be in tension at a global significance of 2.6$\sigma$ with a maximum local significance of 3.4$\sigma$ for the dark matter models they tested.
While more data will be required to confirm or rule out this hint of a discovery, its potential importance motivates investigating theoretical explanations of the event of interest.

Candidates for WIMP dark matter are found
within the context of supersymmetric (SUSY) extensions of the Standard Model (SM) of particle physics.
In our earlier paper \cite{Freese:2026sga}, we proposed Higgsino dark matter as a potential explanation for the LZ event.
The Higgsino is the SUSY partner of the Higgs boson of the SM.
In Ref.~\cite{Freese:2026sga}, we focused on the case of a thermal 1.1 TeV Higgsino. In this paper, we generalize our studies to include the possibility of higher higgsino masses.

The high nuclear recoil energy accessed by LZ, up to $270\ \mathrm{keV}$,
opens up their study to new interactions, beyond the standard Spin Independent (SI) elastic scattering whose nuclear recoil spectrum falls rapidly above $\sim50~\mathrm{keV}$
even for TeV WIMPs.  
Although LZ did previously publish results which went up to these high nuclear recoil energies \cite{LZ:2023lvz},
their new results have signifcantly higher exposure, due to 
220 live days of data used in the current analysis vs. 60 days in their earlier paper.
In the new data, they studied two cases: 1) elastic WIMP-nucleon scattering described by 20 relativistic covariant Lagrangians, in their nonrelativistic limits; and 2) inelastic scattering interactions in which the WIMP transitions into a more massive state.
Both of these possibilities are interesting in the context of high energy recoil spectra (with lower energy components suppressed).
In particular, LZ searched for high-energy nuclear recoils arising due to elastic scattering with all 20 possible interaction types, as well as inelastic interactions arising from spin-independent $(\mathcal{O}_1)$ and spin-dependent $(\mathcal{O}_4)$ operators.\footnote{Here, $\mathcal{O}_1$ and $\mathcal{O}_4$ refer to the Non-Relativistic Effect Field Theory (NREFT) operators that can be used to describe WIMP--nucleon interactions.}

In Ref.~\cite{Freese:2026sga},
we proposed Higgsino dark matter as an explanation of the LZ event of interest, in the context of the inelastic scattering scenario
considered by LZ \cite{LZ:2026axp, Graham:2024syw}.
Higgsinos are split into two mass eigenstates, with a small energy difference $\delta$ between the masses. The lighter Higgsino would be the dark matter. Tree level Z exchange produces inelastic scattering of a Higgsino with a xenon nucleus: the lighter Higgsino can be excited to the higher mass state, giving the xenon nucleus a substantial kick with a large recoil energy.
In our earlier paper~\cite{Freese:2026sga}, we focused on the case of the $\sim1\ \mathrm{TeV}$ Higgsino, as LZ's study had investigated inelastic scattering of 1 TeV dark matter as an explanation of their data; additionally thermal production of $1.1\ \mathrm{TeV}$ Higgsinos in the early  universe produces  the correct relic abundance today to explain the dark matter.
Our paper  showed that predictions for a Higgsino with a mass of $\sim1\ \mathrm{TeV}$, for a mass splitting around $350\ \mathrm{keV}$, that has inelastic spin independent (SI) scattering with the Xenon target, 
would fall within the LZ two-sided confidence interval for their event of interest. 
The following papers proposed the same idea of  Higgsino DM  as explanation of the LZ event on the same day we did: Refs.~\cite{Fan:2026kxx,Wu:2026nhi,Yin:2026jnn} (these papers studied specifically $\sim1.1$ TeV thermal Higgsinos).
 In the current  paper, we generalize our studies to include  higher Higgsino masses, as well as take into account important effects described shortly.
We will show that, although the 1 TeV Higgsino remains only marginally viable in light of bounds from IceCube as discussed below, Higgsinos of mass 2 TeV and above can match data from all existing experiments.

Soon after our earlier paper~\cite{Freese:2026sga}, several important effects came  to light:\\
1) The original calculations only included DM particles from the Standard Halo Model (SHM). The large mass splittings required to interpret the LZ event allow inelastic scattering in xenon only for Higgsinos in the high-velocity tail of the DM distribution. Hence, it is important to include also the effects of the Large Magellanic Cloud (LMC)~\cite{Fan:2026kxx}, which has a mass 10-20\% of the Milky Way.
The LMC speeds up some of the particles in the MW and further contributes DM particles to the high velocity tail.\\
2) Bounds from neutrino observatories: Subsequent to our earlier work~\cite{Freese:2026sga}, Ref.~\cite{Pospelov:2026ewn} immediately pointed out that 
the IceCube neutrino detector places bounds on Higgsino dark matter, due to the lack of detection of neutrinos from the direction of the Sun; see also Ref.~\cite{Nguyen:2026lui}. DM particles traveling through the Sun could be captured, and then annihilate among themselves to produce a detectable neutrino signal. \\
3) Bounds from the LZ sideband: In addition to the LZ event of interest at 248 keV recoil energy, Ref.~\cite{Rodd:2026tyn} took into account the lack of detection by LZ of events at even higher recoil energies, in the sideband at $\sim350-650\ \mathrm{keV}$. However, the LZ collaboration has not published the acceptance in the sideband. LZ has not yet  formally calibrated the nuclear recoil response of the detector up at those high recoil energies, and is now working to do so\footnote{We thank Dan Akerib and Rick Gaitskell for private communication regarding the callibration issue.}.  Hence in this paper we consider results both with and without the empty sideband taken into account.\\
4) Finally, the spin-independent inelastic Higgsino--nucleon scattering cross section is a factor of four larger than previously used (an error pointed out by Ref.~\cite{Pospelov:2026ewn}).

The current paper is an extension of our previous work, in that we take into account all  of these effects as well as generalize our studies to arbitrary Higgsino mass.
We pay particular attention to the IceCube bounds, where we perform a careful analysis that differs substantially from that of many other authors.
We begin with the case of the 1.1 TeV WIMP, which would be thermally produced in the early Universe with the correct relic abundance.
For this case we find (in agreement with Ref.~\cite{Pospelov:2026ewn}) that the combination of the above bounds is difficult to accommodate.
Then we turn to arbitrary Higgsino mass and mass splitting, and examine which regions of parameter space can explain the LZ result as well as accommodate all bounds.
We note that Higgsinos heavier than 1.1 TeV would be overproduced via thermal production in the early Universe, since their contribution to the DM density scales as $m_\chi^2$.
Hence their abundance must be reduced e.g. diluted by entropy production or nonthermal production during a matter dominated phase. 

We will show substantial regions in Higgsino mass and mass splitting ($m_\chi, \delta$) parameter space that accommodate the LZ data as well as all the bounds.
Fig. \ref{fig:higgsino_IceCube_limits} displays the main  results of our work in the $(m_\chi, \delta)$ plane, showing contours that match LZ data as well as IceCube null results, for SHM only as well as SHM+LMC. 
Finally, in the Discussion Section, we conclude with caveats about the remaining uncertainties in many of the calculations leading to predictions, particularly regarding the high velocity tail and the details of the solar physics in the context of IceCube bounds.

The author of Ref.~\cite{Langhoff:2026ujr} previously performed a similar analysis in that he also studied expanded larger Higgsino masses than $1.1\ \mathrm{TeV}$ and took into account the 
issues described above.  However, there are two major differences between his work and ours: First, his treatment of IceCube bounds assumes the extreme scenario where capture--annihilation equilibrium is assumed for all Higgsino masses, a case we also study but deem to be unrealistic beyond 10 TeV Higgsino masses; hence we consider three other scenarios for DM particle cooling inside the Sun.
Secondly, his 68\% C.L. bounds push his results to above 100 TeV; our 90\% C.L. bounds allow masses as low as 2 TeV in agreement with all existing experiments (ignoring the high energy sideband).

Our approach and results differ from those of others in the following ways: (i) we model Higgsino cooling in the Sun without imposing either capture--annihilation equilibrium or thermal equilibrium with the solar core, and assess whether these conditions are reached across the mass range considered, since imposing either equilibrium when it is not achieved overestimates the annihilation rate and yields overly strong IceCube bounds; (ii) we account for the $m_\chi^{-2}$ dependence of the elastic SD Higgsino--nucleon cross section in the large-mass limit~\cite{Hisano:2011cs}, which reduces cooling efficiency and can weaken the IceCube bounds at large Higgsino masses; and (iii) we include the Sommerfeld effect in Higgsino annihilation in the Sun, which can enhance the annihilation rate and strengthen the IceCube bounds, particularly near resonances.

\section{Higgsino dark matter}

Our analysis builds on our Higgsino interpretation introduced in Ref.~\cite{Freese:2026sga}.
Higgsinos are supersymmetric WIMP candidates and provide a well-motivated particle-physics realization of inelastic dark matter~\cite{Krall:2017xij,Co:2021ion,Bhattiprolu:2025zwt,Kowalska:2018toh}.
In the minimal supersymmetric Standard Model (MSSM), after electroweak symmetry breaking, interactions with the Standard Model Higgs split the neutral Dirac Higgsino into two Majorana fermions \cite{Nagata:2014wma},
$\widetilde H_1$ and $\widetilde H_2$, with masses $m_{\widetilde H_{1,2}}\approx\mu \mp \frac{\delta}{2}$, where $\delta$ is the Majorana mass splitting given by
\begin{equation}
    \delta
    \simeq
    m_Z^2
    \left(
        \frac{\sin^2\theta_W}{M_1}
        +
        \frac{\cos^2\theta_W}{M_2}
    \right).
    \label{eq:higgsino_splitting}
\end{equation}
Here $m_Z$, $M_1$, and $M_2$ are the Z boson, bino and wino masses, respectively, and $\theta_W$ denotes the Weinberg angle.
In the limit of heavy gauginos ($M_{1,2}\gg m_Z,\mu$), the two neutral Higgsinos constitute a narrowly split pseudo-Dirac pair, $\delta\ll\mu$.
A spectrum in which the gauginos and other superpartners are much heavier than the weak scale is compatible with the absence of new physics at the LHC 
\cite{Arkani-Hamed:2004ymt,Giudice:2004tc,Gonski:2025wzh}.

For an illustration of the heavy-gaugino limit, we take ($M_1=M_2=M>0$), for which Eq.~\eqref{eq:higgsino_splitting} gives $M\simeq m_Z^2/\delta\simeq2.4\times10^7\ \mathrm{GeV}\;(350\ \mathrm{keV}/\delta)$. Such heavy electroweak gauginos lie far above conventional naturalness expectations and entail an electroweak fine-tuning in the MSSM, leaving a residual little hierarchy problem \cite{Baer:2015rja}. While this fine-tuning does not preclude the Higgsino interpretation of the LZ event, it represents a theoretical cost of the required supersymmetric spectrum.

If the lighter Higgsino $\widetilde H_1$ is the lightest supersymmetric particle, it can be stable and serve as a WIMP dark-matter candidate \cite{Ellis:1983ew}.
On the contrary, the heavier state $\widetilde H_2$ decays on a timescale much shorter than the age of the Universe, leaving $\widetilde H_1$ as the surviving dark-matter component.

In the nearly pure Higgsino limit, the dominant neutral-current coupling connects $\widetilde H_1$ and $\widetilde H_2$.
Consequently, tree-level $Z$ exchange produces the inelastic transition from $\widetilde H_1$ to $\widetilde H_2$. 
The mass of the initial and the final particle state differs by $m_{\widetilde{H}_2}-m_{\widetilde{H}_1}=\delta$
\cite{Tucker-Smith:2001myb}.
The resulting spin-independent interaction is coherent and couples primarily to neutrons~\cite{Nagata:2014wma,Kowalska:2018toh,Essig:2007az}. 
The corresponding Higgsino--nucleon cross section (for vector coupling) is 
\begin{equation}
    \sigma_{\widetilde H n}
    =
    \frac{G_F^2\mu_n^2}{2\pi}
    \simeq
    7.43\times10^{-39}~{\rm cm}^2 ,
    \label{eq:higgsino_neutron_xsec}
\end{equation}
where $\mu_n$ is the Higgsino--nucleon reduced mass.
Although this cross section greatly exceeds the elastic cross sections tested by conventional direct-detection searches, the excitation threshold can strongly suppress the observable scattering rate.
Increasing $\delta$ restricts the incident population capable of producing the heavier state, making terrestrial detection particularly sensitive to the high-speed tail of the local dark-matter distribution.

Elastic scattering is also possible through Higgs exchange and electroweak loop processes, but these contributions are much smaller.
The tree-level Higgs coupling requires a gaugino admixture and is suppressed by $m_Z/M_{1,2}$ in the heavy-gaugino limit~\cite{Krall:2017xij}.
The loop-induced spin-independent amplitude is further reduced by cancellations between scalar and spin-2 contributions~\cite{Hisano:2011cs,Hisano:2012wm,Chen:2019gtm}, which can place the predicted signal below conventional estimates of the neutrino floor. 
Although these suppressed elastic interactions do not account for the high-energy LZ event, they can play an important role in cooling captured Higgsinos in the Sun, thereby affecting the IceCube bounds.

\section{Our likelihood Analysis}

In this section we describe our  likelihood analysis. We construct an approximate binned likelihood from publicly available information~\cite{LZ:2026axp}.
This approach accounts for changes in the predicted signal distribution across bins, improving on the total-rate rescaling used in our earlier estimate.
The Poisson log likelihood is
\begin{equation}
\begin{aligned}
\ln\mathcal{L}(\sigma,m_\chi,\delta)=\sum_{i=1}^{96}
\bigg\{&n_i\ln\left[b_i+(\sigma/\sigma_{\rm ref})\; s_i(m_\chi,\delta)\right]
\\[1mm]
    &
-\left[b_i+(\sigma/\sigma_{\rm ref})\; s_i(m_\chi,\delta)\right]\bigg\},
\end{aligned}
\label{eq:approximate_likelihood}
\end{equation}
where $m_\chi$ is the Higgsino DM mass, $n_i$, $b_i$, and $s_i(m_\chi,\delta)$ are, respectively, the observed count, fixed background expectation, and reference DM signal (evaluated at a reference DM-nucleon cross section $\sigma_{\rm ref}=10^{-39}~\mathrm{cm}^{2}$) in bin $i$, while the DM-nucleon cross section $\sigma$ rescales the signal. 
We ignore constant factors in the expression of the log likelihood, which cancel in the likelihood ratio we are ultimately interested in. 
We adopt the binning\footnote{Xenon detectors have two detector-response observables: $S1c$ and $S2c$ , the position-corrected prompt scintillation light and delayed electroluminescence light, respectively, measured in photons detected. We bin in three $S1c$ ranges ($S1c<250$, $250<S1c<500$, and $500<S1c<600$), each divided into 32 bins of width $0.5$ in
$z_{\rm NR}\equiv[\log_{10}(S2c)-\mu_{\rm NR}]/\sigma_{\rm NR}$, giving 96 bins in total. 
The quantity $\mu_{\rm NR}$ denotes the central value of $\log_{10}(S2c)$ for nuclear recoils at fixed $S1c$, while $\sigma_{\rm NR}$ denotes the corresponding band width.}
of Fig.~5 of the LZ paper~\cite{LZ:2026axp}.
We calculate the differential nuclear-recoil rate $dR/dE_R$, expressed per unit detector mass, time, and recoil energy $E_R$, using \textsc{WimPyDD}~\cite{Jeong:2021bpl}. 
Expected event counts are obtained for the LZ exposure of $2.84~\text{tonne-year}$ by folding the recoil spectrum with the reported detection efficiency.\footnote{We assign each $E_R$ to an $S1c$ value using the approximate one-to-one correspondence inferred from Fig.~2 of Ref.~\cite{LZ:2026axp} and model $x\equiv\log_{10}(S2c)$ at fixed $S1c$ as Gaussian with mean $\mu_{\rm NR}$ and standard deviation $\sigma_{\rm NR}$. Because the bins are defined in the standardized variable $z_{\rm NR}$, the resulting Gaussian bin probabilities depend only on the bin edges; the explicit values of $\mu_{\rm NR}$ and $\sigma_{\rm NR}$ therefore cancel within this approximation.}

\begin{figure}[t]
    \centering

    \includegraphics[width=\columnwidth]
    {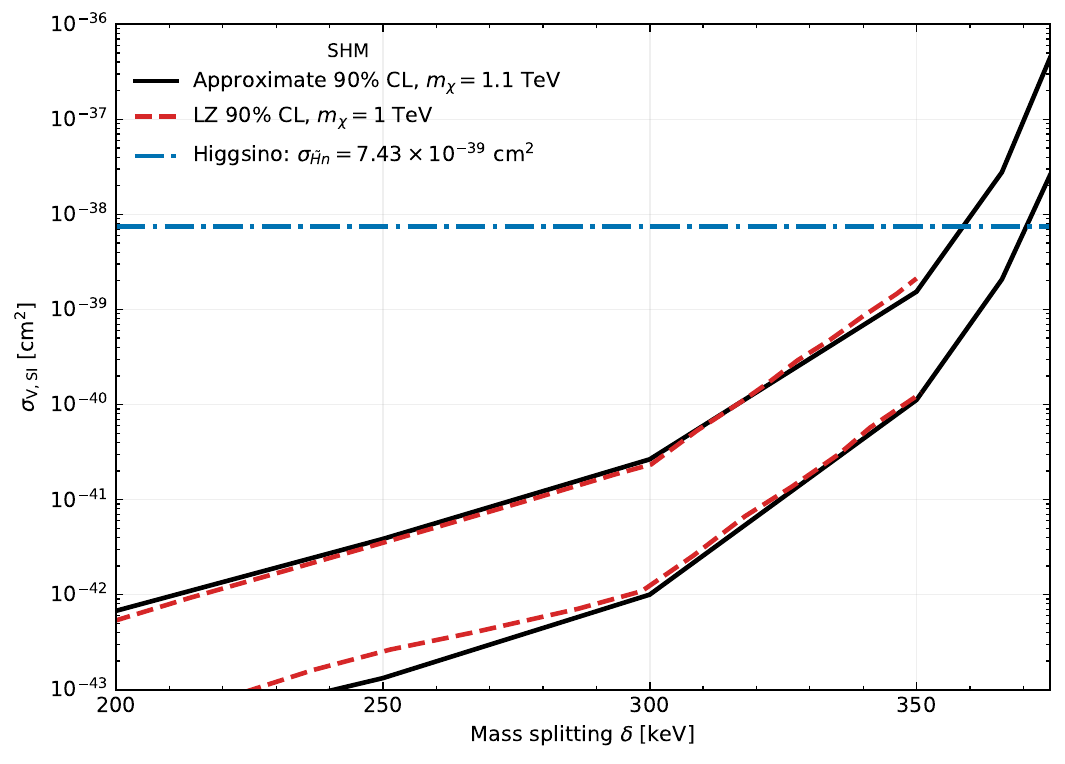}

    \vspace{0.1ex}


    \includegraphics[width=\columnwidth]
    {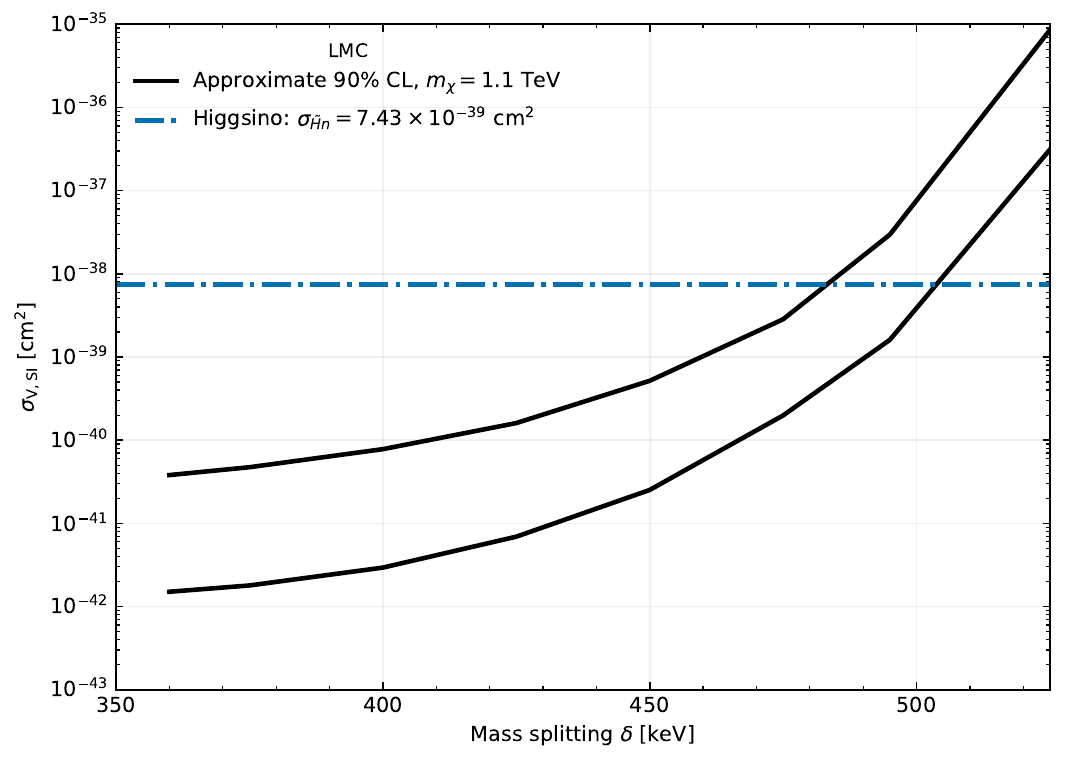}

    \caption{Illustration of the effect of the Large Magellenic Cloud's enhancement of the high velocity tail of the DM distribution:
    The case of the 1.1 TeV thermal Higgsino and LZ data only, without (upper panel)  and with (lower panel) the effects of the LMC.
    Comparison of Higgsino--nucleon SI inelastic scattering cross section with the two-sided 90\% confidence level (CL)  intervals inferred from LZ data
    at $m_\chi=1.1~\mathrm{TeV}$, as a function of higgsino mass splitting $\delta$.
    In both panels,  black solid curves indicate the
    two-sided 90\% confidence interval inferred using our approximate likelihood. 
    The horizontal blue dot-dashed line denotes the  Higgsino--nucleon cross section.
    \textit{Upper panel:} Results assuming the Standard Halo Model (SHM)~\cite{Baxter:2021pqo}. The red dashed curves show two-sided 90\% confidence interval for $m_\chi=1~\mathrm{TeV}$
    found by the LZ collaboration, which they obtained only up to $\delta=350~\mathrm{keV}$~\cite{LZ:2026axp} (n.b. note the agreement between the red and black curves). 
    \textit{Lower panel:} Results including the enhancement of the high-speed tail of the local DM distribution induced by the Large Magellanic Cloud (LMC)~\cite{Smith-Orlik:2023kyl}. 
}

    \label{fig:higgsino_limits}
\end{figure}
We digitize the observed counts (science sample) and total post-fit background from Fig.~5 of Ref.~\cite{LZ:2026axp} and hold the background template fixed. This template corresponds to the post-fit LZ model obtained under the $m_\chi=1~\mathrm{TeV}$ elastic $\mathcal{L}_{10}^{s}$ signal hypothesis.\footnote{$\mathcal{L}_{10}^s$ is one of the 20 interaction Lagrangians presented in Ref.~\cite{Anand:2013yka}.} 
Holding it fixed introduces negligible mismodeling for other signal
hypotheses. The post-fit normalizations of Ref.~\cite{LZ:2026axp} lie within
$0.35\sigma$ of the pre-fit expectations for every component above one
event, the fitted signal absorbs only $1.0^{+1.4}_{-0.7}$ of $1713\pm39$
total counts, and the post-fit uncertainty on the total background is
$2.3\%$. In the highest-$S1c$ range, where the event of interest lies, the integrated background
is $0.0106\pm0.0008$ counts, so any re-profiling under a different signal
hypothesis shifts the expectation there by $\lesssim10^{-3}$ events,
which is negligible compared with the single observed event.

We extend the analysis presented in Fig.~1 of our earlier paper~\cite{Freese:2026sga} to mass splittings above $350~\mathrm{keV}$ to confirm our earlier claim that the thermal Higgsino model is compatible with the LZ event, using the approximate likelihood described above. 
At this stage, we omit the IceCube constraints and the null result in the high-energy LZ sideband, both of which are examined in detail later in this paper. 
Our statistical inference is based on the likelihood-ratio statistic
\begin{equation}
q_\sigma=-2\ln
\frac{\mathcal{L}(\sigma;m_\chi,\delta)}
{\mathcal{L}(\widehat{\sigma};m_\chi,\delta)},
~
\widehat{\sigma}
=\underset{\sigma\geq0}{\operatorname{argmax}}\;
\mathcal{L}(\sigma;m_\chi,\delta),
\label{eq:likelihood_ratio}
\end{equation}
whose distribution is calibrated using $30{,}000$ Poisson Monte Carlo toy pseudoexperiments at each tested point.
Figure~\ref{fig:higgsino_limits} shows the results for the thermal Higgsino assuming either the Standard Halo Model~\cite{Baxter:2021pqo} or a velocity distribution that includes the enhancement of the high-speed tail due to the Large Magellanic Cloud~\cite{Smith-Orlik:2023kyl}, as discussed in the next section.

Our likelihood disfavors DM models that predict an appreciable additional signal at recoil energies below $\sim200~\mathrm{keV}$, where the LZ data are well described by the reported background model.
It favors models predicting approximately one signal event in the recoil-energy interval $200$--$270~\mathrm{keV}$, which contains the event of interest.

\section{Effects of the Large Magellenic Cloud on the High Velocity Tail of the DM Distribution}
\label{sec:LMC}

The LZ analysis adopted the standard astrophysical parameters recommended in Ref.~\cite{Baxter:2021pqo}, including the Standard Halo Model (SHM) velocity distribution. 
However, high energy recoil events can be very sensitive to the high-speed tail of the local DM distribution, which is not well-known.
The interaction of our dark matter halo with neighbor halos, such as the Large Magellanic Cloud (LMC)~\cite{Besla:2019xbx,Smith-Orlik:2023kyl}, can accelerate the DM particles in our halo and allow high-speed DM particles from neighboring halos to enter ours as well as gravitationally significantly enhance the high-speed tail of the local DM distribution.
Including the effects of the  LMC is particularly important as it weighs $\sim 10-20\%$ of the Milky Way.
In our earlier paper~\cite{Freese:2026sga}, we took into account only the WIMPs from the SHM.  

In our current analysis
 we explore two cases, a SHM only case~\cite{Baxter:2021pqo}, and a SHM-plus-LMC case with the time-averaged, present-day MW+LMC halo velocity distribution shown in Fig.~11 of Ref.~\cite{Smith-Orlik:2023kyl}.
Because the figure provides no information beyond $v_{\min}\simeq950~\mathrm{km/s}$, we conservatively set $\eta(v_{\min})=0$ above this value.

In this Section, we explore the effects of including the LMC for the case of the thermal 1.1 TeV Higgsino, as an extension of our earlier work.
In subsequent sections, we generalize to arbitrary Higgsino masses, with and without the inclusion of LMC effects.

Figure~\ref{fig:higgsino_limits} shows the results in the $(\sigma, \delta)$ plane for the $1.1\ \mathrm{TeV}$ Higgsino case for the SHM only (upper panel) and including the effects of the LMC (lower panel).
The red curves in the upper panel are the two-sided 90\% confidence intervals presented by the LZ collaboration for the SHM only, 
and black curves in both panels are those inferred from our approximate likelihood for the LZ data.   Note that LZ presented results for the SHM alone at $m_\chi=1\ \mathrm{TeV}$ and only for mass splittings $\delta\le350\ \mathrm{keV}$.  The horizontal dot-dashed blue lines illustrate the inelastic Higgsino scattering prediction for the cross section.
The cross sections are expressed in the vector-coupling normalization $\sigma_{V,\mathrm{SI}}$ instead of the scalar normalization $\sigma_{S,\mathrm{SI}}$, using the LZ conversion $\sigma_{V,\mathrm{SI}}\simeq3.2\;\sigma_{\mathrm{S,SI}}$ for xenon~\cite{LZ:2026axp}.


In our earlier paper~\cite{Freese:2026sga}, we analyzed the 1.1 TeV Higgsino using the SHM velocity distribution only.  We found agreement with the LZ collaboration's 90\% confidence intervals presented in \cite{LZ:2026axp} for the 248 keV event for mass splitting $\delta \sim 350$ keV.  This result is shown in the upper panel of Fig.~\ref{fig:higgsino_limits}.
Running our likelihood in that same configuration reproduces their published interval near $\delta=350\ \mathrm{keV}$. 
This motivates the use of our likelihood as we extend our analysis to larger Higgsino masses and mass splittings.

Further we must take into account the fact that the cross section for the 1.1 TeV Higgsino is a factor of four higher than what we used in Ref.~\cite{Freese:2026sga} (see Ref.~\cite{Pospelov:2026ewn}).
Using the likelihood analysis presented in this paper, we find that, the predicted Higgsino cross section is compatible with the LZ data at a mass splitting slightly above $350~\mathrm{keV}$, as suggested in our earlier work for the SHM alone. 
Specifically, for the SHM only (upper panel in Fig.~\ref{fig:higgsino_limits}), for the $1.1\ \mathrm{TeV}$ Higgsino, we find that the Higgsino--nucleon cross section lies within the inferred interval for $360~\mathrm{keV}\lesssim\delta\lesssim370~\mathrm{keV}$.
Including the particles in the SHM + LMC  velocity distribution (lower panel) shifts the compatible range to $485~\mathrm{keV}\lesssim\delta\lesssim505~\mathrm{keV}$, approximately $130~\mathrm{keV}$ above the SHM range, highlighting the impact of the enhanced high-speed tail on the result.

Clearly a proper treatment of the WIMP distribution in the Galaxy, with attention to the particles near escape velocity, is essential.   Many uncertainties remain.
Studies (including ours) of the effects of the LMC on the high velocity tail of the dark matter in the MW
have used the results of simulations of an LMC analogue known as ``halo-13" in Ref.~\cite{Smith-Orlik:2023kyl}.  However, this halo has an extremely large ratio of LMC to MW analogue mass, $M_{\rm infall}^{\rm LMC}/M_{200}^{\rm MW} = 0.27$ (at the time of closest approach i.e. pericenter of the LMC at $\sim 133$Myr before the current day), where $M_{\rm infall}^{\rm LMC}$ is the LMC analogues's virial mass at infall and $M_{200}^{\rm MW}$ is the virial mass of the MW analogue.  If this value is an overestimate, then the enhancement to the MW's high velocity tail may also be an overestimate. 
On the other hand, the high velocity tail may also be perturbed (and possibly enhanced) by other bodies in/near the Milky Way, including the Small Magellenic Cloud and dwarf galaxies \footnote{The LMC, SMC, and at least 8 dwarfs \cite{bechtol2015eight}  are all in the same part of the sky and on similar orbits. Hence it's plausible they all came into the Milky Way together.  Their group mass may be substantial and hence further modify the high velocity tail of interest.}.  Hence it is hard to correctly ascertain the high velocity tail of the DM distribution, and consequently to make detailed predictions of DM events in detectors including LZ. These uncertainties could in principle enlarge the 90\% confidence regions we have presented here, e.g. potentially allowing the thermal 1.1 TeV Higgsino  better agreement with the combination of LZ data and IceCube bounds. 

In the next Sections, we take into account additional bounds on the Higgsino interpretation of the LZ event
due to the empty LZ sideband as well as IceCube bounds due to null results in the direction of the Sun.
These bounds severely constrain the interpretation of the LZ result as due to thermal $1.1\ \mathrm{TeV}$ Higgsinos and instead favor higher Higgsino masses.
Hence we will consider arbitrary Higgsino masses and mass splittings, for both cases of SHM dark matter only as well as SHM + LMC as discussed in this Section.

\section{High-energy sideband in LZ Data}
Rodd et al. suggested that any explanation of the LZ event at $\delta = 350\ \mathrm{keV}$ should also be consistent with the lack of events in the LZ experiment at higher energies, in its ``sideband" at recoil energies $\sim 350$--$650~\mathrm{keV}$~\cite{Rodd:2026tyn}.
However, the LZ collaboration has not published the acceptance in this sideband and did not include it
in the DM signal likelihood.
The nuclear recoil response of the detector up at those high recoil energies has not yet been calibrated and is the subject of current work by the LZ collaboration.  Hence in this paper we will present results both with and without the empty sideband taken
into account.

In those cases where we do require compatibilty of our model with an empty sideband,
we extend our LZ likelihood to include the sideband.  We can then identify regions in the $(m_\chi,\delta)$ plane compatible with the LZ region of interest (ROI) as well as the empty sideband.
We extend our 96-bin likelihood by a Poisson factor $\mathcal{L}_{\rm SB}=\exp(-N_{\rm SB})$, modeling the sideband as $350<E_R<650~\mathrm{keV}$ with zero observed events, zero expected background, and unit signal acceptance.
Here $N_{\rm SB}$ is the predicted differential recoil rate integrated over this interval and multiplied by the exposure.
Because the acceptance for DM-induced recoils in the experimental sideband has not been published, the resulting constraint is conditional on our acceptance and energy-window approximations~\cite{Rodd:2026tyn}.

Fig. \ref{fig:higgsino_IceCube_limits} illustrates the effect of the sideband on the 
LZ confidence regions  at 90\% CL in the Higgsino parameter space.
    The upper (lower) panel uses the SHM (SHM + LMC) velocity distribution for the LZ prediction.
    Orange dashed and blue solid contours show the approximate joint 90\% confidence regions in $(m_\chi,\delta)$ obtained from our likelihood analysis with and without the high-energy sideband, respectively, at fixed Higgsino coupling.
    The star and diamond mark the corresponding best-fit points, respectively. 
    We note that there are two regions compatible with empty sideband events. One region corresponds to  lower DM mass $m_{\chi} < 1$ TeV, in which case very few high energy recoil events are produced; but this low mass region is ruled out by IceCube neutrino bounds described in the next Section. The second region corresponds to higher Higgsino mass, and it is compatible with the empty sideband because the particle abundance and hence expected count rate decreases $\sim 1/m_{\chi}$.

    In the rest of the paper we will present results both with and without the empty sideband taken into account.


\section{Solar Bounds from Neutrino Telescopes}\label{sec:icecube}

\begin{figure}[t]
    \centering
    \includegraphics[width=\columnwidth]
    {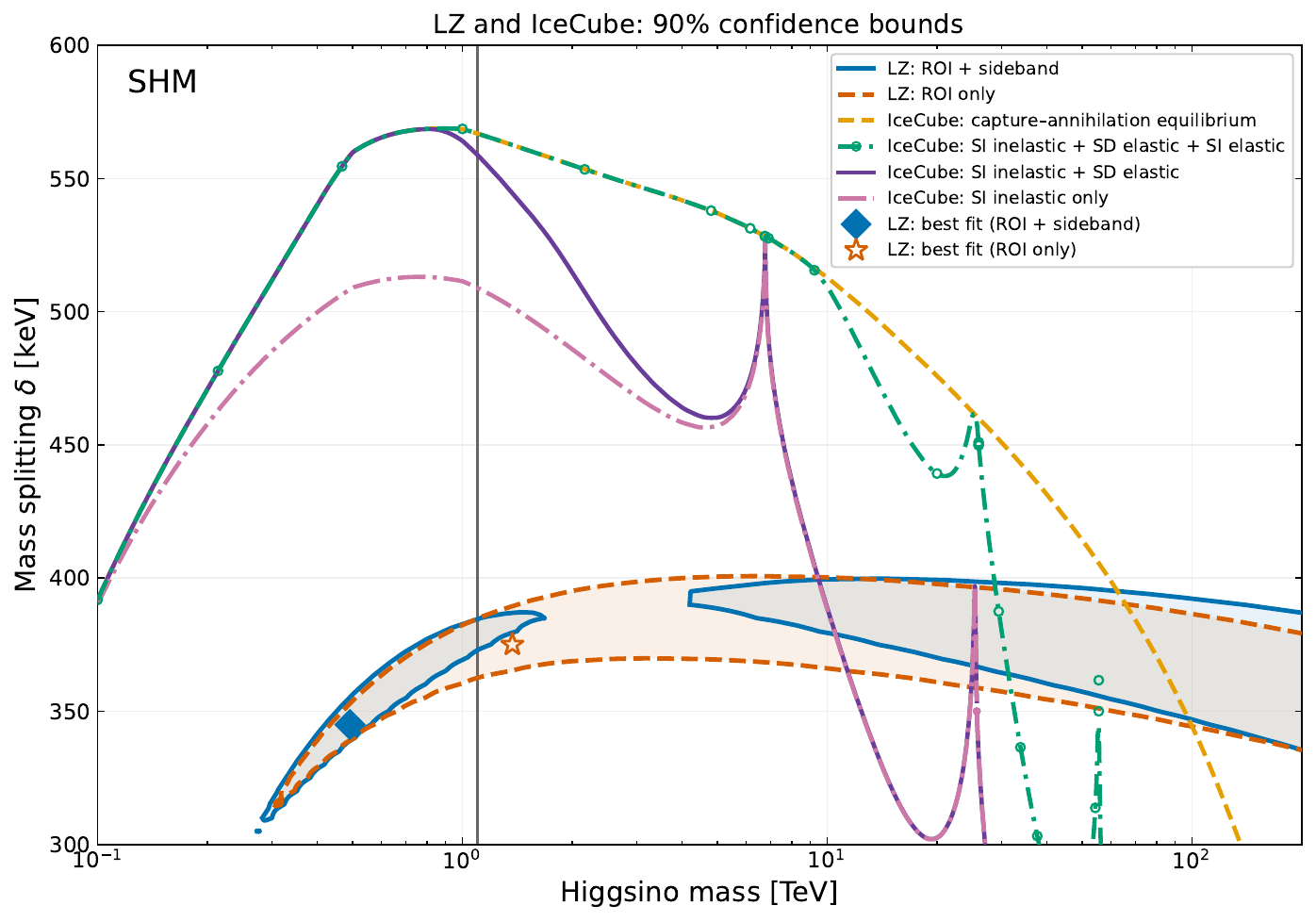}

    \vspace{0.1ex}

    \includegraphics[width=\columnwidth]
    {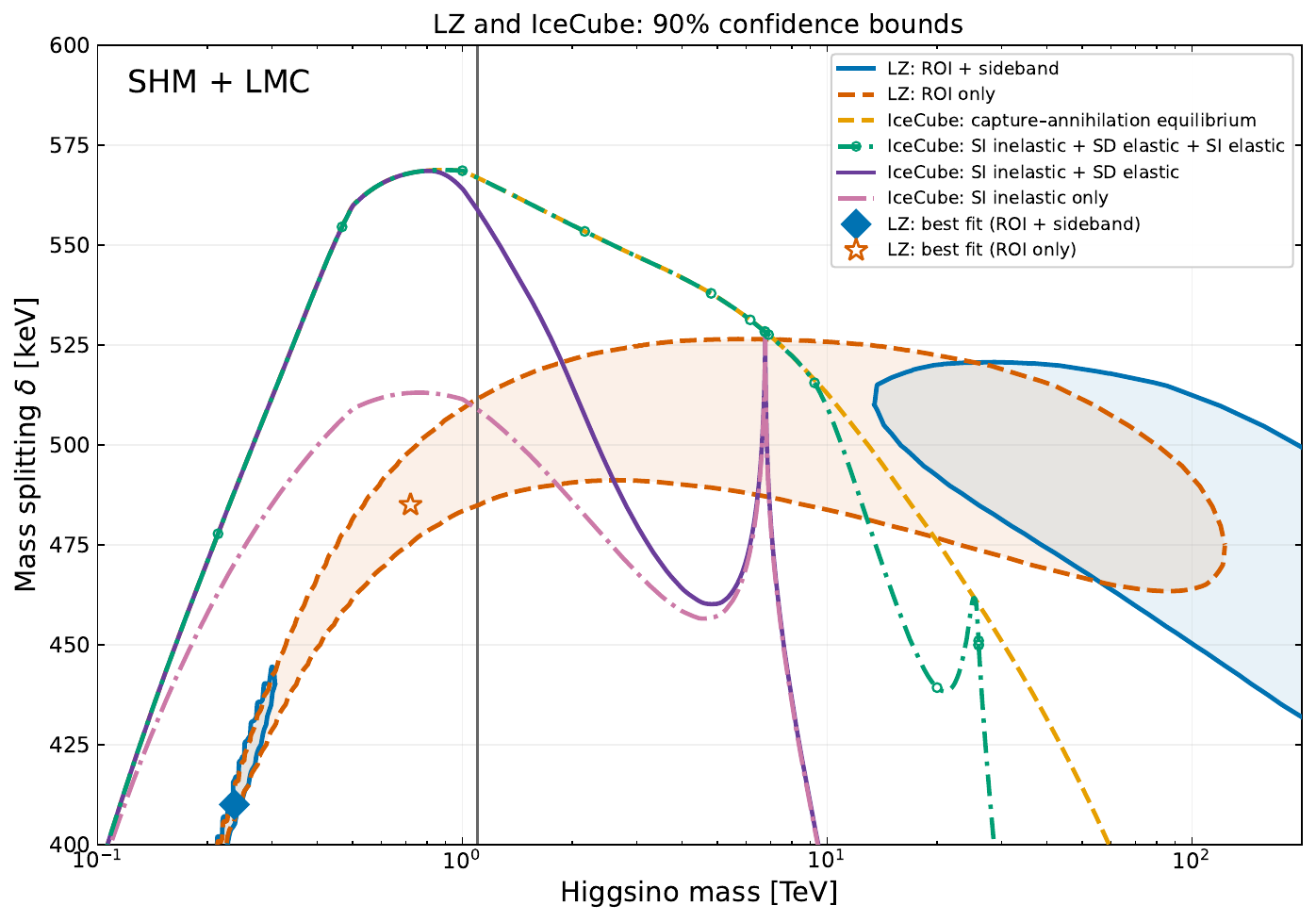}

    \caption{LZ confidence regions and estimated IceCube limits at 90\% CL in the Higgsino parameter space.
    The upper (lower) panel uses the SHM (SHM + LMC) velocity distribution for the LZ prediction.
    Orange dashed and blue solid contours show the approximate joint 90\% confidence regions in $(m_\chi,\delta)$ obtained from the ROI likelihood and its extension including the high-energy sideband, respectively, at fixed Higgsino coupling.
    The star and diamond mark the corresponding best-fit points, respectively.
    IceCube lower limits on $\delta$ are shown for capture--annihilation equilibrium (gold dashed), inelastic SI cooling alone (magenta dot-dashed), additional elastic SD and SI cooling with $\sigma_{\mathrm{SI,elastic}}=4\times10^{-50}~\mathrm{cm^2}$ (green with circles), and additional elastic SD cooling with $\sigma_{\mathrm{SI,elastic}}=0$ (purple solid).
    Both elastic SI cross sections considered here are consistent with the theoretical uncertainty band~\cite{Chen:2019gtm}.
    The cooling-dependent limits use the stationary-nucleus and uniform-DM-density approximations described in the text.
    We adopt IceCube's published annihilation-rate limits for the $W^+W^-$ channel~\cite{IceCube:2025fcu} for $m_\chi\leq10\ \mathrm{TeV}$, while above $10~\mathrm{TeV}$, the adopted limits are held fixed at its $10~\mathrm{TeV}$ value.
    The vertical gray line marks the thermal Higgsino mass, $m_\chi=1.1~\mathrm{TeV}$.}

    \label{fig:higgsino_IceCube_limits}
\end{figure}

Neutrino telescopes provide complementary probes of dark matter particles through searches for neutrinos from DM annihilation in the Sun\footnote{Silk, Srednicki, and Olive~\cite{Silk:1985ax} first proposed searches for neutrinos from DM annihilation in the Sun; Freese~\cite{Freese:1985qw} and Krauss, Srednicki, and Wilczek~\cite{Krauss:1985aaa} first proposed searches for neutrinos from DM annihilation in the Earth.}.
In particular, the absence of a neutrino excess from the direction of the Sun in IceCube and Super-Kamiokande places strong constraints on the DM particles studied in this paper.
The same inelastic DM--nucleus scattering process that could explain the LZ event can also occur as DM traverses the Sun.
Inelastic interactions with solar nuclei cause DM particles to lose energy, allowing some to become gravitationally bound to the Sun when their speeds fall below the local solar escape speed.
Because the solar gravitational field accelerates incident DM particles, inelastic scattering on heavy solar nuclei can occur at mass splittings beyond the reach of terrestrial xenon experiments.
Once captured, DM particles can cool by losing energy through repeated scattering with solar nuclei.
Inelastic cooling stops when scattering becomes kinematically inaccessible, while elastic interactions can allow further energy loss.
This cooling concentrates the captured population closer to the solar center, increasing its density and enhancing the annihilation rate.
If the Higgsino interactions are sufficiently effective, cooling can proceed rapidly enough for Higgsino capture and annihilation to approach equilibrium.
Capture--annihilation equilibrium implies that the number of Higgsinos captured per unit time equals the number lost to annihilation, yielding the maximum annihilation rate sustained by a capture rate.
Understanding when cooling becomes inefficient or ceases is essential, as well as  identifying whether or not capture--annihilation equilibrium is reached. 
DM annihilation in the Sun produces neutrinos that can be detected by neutrino telescopes.
The absence of a DM-induced solar neutrino excess in IceCube can therefore constrain Higgsino properties, specifically the particle mass and mass splitting~\cite{Pospelov:2026ewn,Nguyen:2026lui}.
Here we estimate these constraints under different assumptions about the interactions responsible for cooling the captured population, and present our results in Fig.~\ref{fig:higgsino_IceCube_limits}.

\subsection{Higgsino capture in the Sun}

As Higgsinos travel through the Sun, they scatter inelastically with the nuclei in the Sun, lose energy due to the interactions, and some become gravitationally bound to the Sun. 
To estimate the Higgsino capture rate in the Sun, we use the standard DM capture formalism~\cite{Press:1985ug}, together with the treatment of inelastic Higgsino scattering described in Refs.~\cite{Pospelov:2026ewn,Nguyen:2026lui}. 

The gravitational field of the Sun accelerates DM particles of incident velocity $u$ in the Solar frame to speeds $w(r,u)=\sqrt{u^2+v_{\rm esc}^2(r)}$ in the Sun's rest frame, as follows from energy conservation. Here, 
$v_{\rm esc}(r)$ is the local solar escape speed, calculated from the gravitational potential corresponding to the total solar density profile of the BS05(AGS,OP) solar model~\cite{Bahcall:2004pz}.
We work in the stationary-nucleus approximation, neglecting the thermal motion of nuclei in the Sun.
In this approximation, the DM speed after acceleration $w$ is the DM–nucleus relative speed at the collision. 

Capture is dominated by the same inelastic SI interaction that could account for the LZ event; we include only this interaction in the capture calculation and treat the loop-induced elastic interactions as contributing to subsequent cooling~\cite{Pospelov:2026ewn}.
The differential cross section for inelastic SI Higgsino scattering on a nucleus $A$ is~\cite{Essig:2007az,Goodman:1984dc}
\begin{equation}
\frac{d\sigma_{\chi A}}{dE_R}
=\frac{G_F^2m_A}{4\pi w^2}\left[N_A-(1-4\sin^2\theta_W)Z_A\right]^2 F_A^2(q),
\end{equation}
where $q=\sqrt{2m_AE_R}$ is the magnitude of the momentum transfer from the nucleus to the DM particle, $Z_A$ and $N_A$ are the proton and neutron numbers, $m_A$ is the nuclear mass, and $F_A(q)$ is the nuclear form factor, for which we adopt the Helm form.
Energy and momentum conservation determine the maximum $E_R^+$ and minimum $E_R^-$ nuclear recoil energies kinematically allowed in inelastic scattering:
\begin{equation}
    E_R^\pm=\frac{\mu_A^2}{2m_A}
\left(w\pm\sqrt{w^2-\frac{2\delta}{\mu_A}}\right)^2,
\end{equation}
where $\mu_A$ is the Higgsino--nucleus reduced mass for nuclear species $A$.
We calculate the Higgsino DM capture rate assuming that a DM particle can interact only once before being captured or escaping~\cite{Nguyen:2026lui}. 
In this single-scattering approximation, a Higgsino must lose at least $m_\chi u^2/2$ of kinetic energy to become captured, where $m_\chi$ is the Higgsino DM mass.
The minimum nuclear recoil energy required for Higgsino capture is therefore $E_R^{\mathrm{cap}}=m_\chi u^2/2-\delta$.

We calculate the capture rate $C$ for Higgsinos traversing the Sun by summing the inelastic SI scattering rate over the nuclear species in our adopted solar profile \cite{Bahcall:2004pz,Asplund_2021} and integrating over radius, incident speed, and nuclear recoil energy:
\begin{equation}
\begin{aligned}
C=\frac{\rho_\chi}{m_\chi}
\sum_A\int_0^{R_\odot}4\pi r^2\;dr\;n_A(r)
&\times\int_0^{u_{\rm max}}du\;\frac{F_\odot(u)}{u}\;w^2
\\
&\times\int_{E_R^{\rm min}}^{E_R^+}dE_R\;
\frac{d\sigma_{\chi A}}{dE_R},
\end{aligned}
\end{equation}
where $n_A(r)$ is the nuclear number density, $E_R^{\mathrm{min}}=\max\left[E_R^-,E_R^{\mathrm{cap}}\right]$, and $u_{\rm max}$ is the largest asymptotic speed for which capture on species A at radius r is kinematically allowed. 
For the velocity distribution we assume the SHM with $F_\odot(u)=u^2\int d\Omega\;f_{\rm Gal}(\mathbf u+\mathbf v_\odot)$, where $f_{\rm Gal}(\mathbf v)\propto e^{-v^2/v_0^2}\Theta(v_{\rm esc}^{\rm Gal}-v)$ is normalized to unity, and $v_0=220~\mathrm{km\;s^{-1}}$, $v_\odot=232~\mathrm{km\;s^{-1}}$, and $v_{\rm esc}^{\rm Gal}=544~\mathrm{km\;s^{-1}}$.
Because solar gravity supplies much of the collision speed, capture generally depends less strongly on the halo's high-speed tail than terrestrial xenon scattering, although sensitivity remains near kinematic thresholds~\cite{Nguyen:2026lui}.  Thus the solar bounds from neutrino observatories remain unchanged by adding the effects of the LMC,
and we may use
the same velocity distribution above when plotting neutrino bounds in both panels of Fig.~\ref{fig:higgsino_IceCube_limits}.
We assume that Higgsinos constitute all of the local DM, with $\rho_\chi=0.3~\mathrm{GeV\;cm^{-3}}$.

We calculate the number density of each of the 51 target nuclear species in our adopted solar profile as $n_A(r)=X_A(r)\rho_\odot(r)/m_A$. 
We adopt the total solar density $\rho_\odot(r)$ and the radial mass fractions $X_A(r)$ of $^1$H, $^3$He, $^4$He, $^{12}$C, $^{14}$N, and $^{16}$O from the BS05(AGS,OP) solar model \cite{Bahcall:2004pz}. 
For the remaining 45 species, we assume constant mass fractions calculated by the elemental abundances in Ref.~\cite{Asplund_2021}, assuming
that heavy elements have the same density profile up to an overall normalization, as considered in Ref.~\cite{Pospelov:2026ewn}.

Inelastic Higgsino--nucleus scattering requires $\delta\leq\mu_Aw^2/2$.
Although hydrogen is the most abundant solar element, inelastic scattering at splittings of several hundred keV requires heavier targets, such as iron, nickel, lead, and uranium, with the accessible species depending on the Higgsino mass and splitting~\cite{Pospelov:2026ewn,Nguyen:2026lui}.
The capture rate $C$ is used below to calculate the Higgsino annihilation rate and the characteristic capture--annihilation equilibrium timescale $t_\mathrm{eq}$.

\subsection{DM distribution in the Sun: cooling and possible thermalization}

To compute the DM annihilation rate in the Sun, we need the spatial distribution of the captured DM population.
Captured Higgsinos can lose energy through repeated scattering on solar nuclei, sinking towards the center of the Sun.
The key question in this subsection is the spatial distribution of DM after cooling.
We approximate the resulting DM population as uniformly distributed within a sphere of radius $R$, which we determine in the remainder of this subsection.

A bound Higgsino traverses the Sun repeatedly, increasing the probability of undergoing multiple scatterings in the solar interior before annihilating.
Several scattering channels contribute to this cooling.
First, Higgsinos can continue to scatter through the inelastic SI interaction after capture.
Cooling through this channel stops when Higgsino energies fall below the kinematic threshold for inelastic scattering, even on the heaviest solar nuclei.
As a Higgsino loses energy, the last species in our solar composition on which it can scatter inelastically is uranium.
Following Ref.~\cite{Pospelov:2026ewn}, we approximate the trajectories of captured Higgsinos as radial when calculating their subsequent cooling.
The threshold for scattering on uranium determines the stall turning point $R_{\mathrm{stall}}$:
\begin{equation}
    v_{\mathrm{esc}}^2(0)-v_{\mathrm{esc}}^2(R_{\mathrm{stall}})
    =\frac{2\delta}{\mu_{\text{U}}}\,,
\end{equation}
where $v_{\mathrm{esc}}(R)$ is the local solar escape speed and $\mu_{\text{U}}$ is the Higgsino--uranium reduced mass.
Following the simplified treatment of Ref.~\cite{Pospelov:2026ewn}, we approximate the preceding inelastic cooling as instantaneous and initialize the elastic cooling evolution at this stall radius.
We treat solar nuclei as stationary throughout the scattering calculation.

Cooling can continue through elastic Higgsino scattering on solar nuclei, which has no inelastic excitation threshold.
Elastic Higgsino interactions are loop suppressed and much weaker than the inelastic interaction relevant to LZ, but they can play a crucial role in cooling Higgsinos in the Sun.
For the elastic SD Higgsino--proton interaction, we use the loop-induced mass dependence of Ref.~\cite{Hisano:2011cs}, normalized to $5\times10^{-47}~\mathrm{cm}^2$ at $m_\chi=1.08~\mathrm{TeV}$.
Its heavy-mass behavior is approximately $\sigma_{\mathrm{SD,elastic}}\sim5\times10^{-47}~\mathrm{cm}^2\times(1.08~\mathrm{TeV}/m_\chi)^2$, although the full loop function is used in the calculation.
The loop-induced elastic SI Higgsino--nucleon amplitude exhibits a strong cancellation between scalar and spin-2 contributions, suppressing the cross section and enhancing the relative theoretical uncertainty in its prediction~\cite{Chen:2019gtm}.
We consider two benchmark values for the elastic SI nucleon cross section, $\sigma_{\mathrm{SI,elastic}}=4\times10^{-50}~\mathrm{cm}^2$ and $\sigma_{\mathrm{SI,elastic}}=0$, both consistent with the uncertainty range in Ref.~\cite{Chen:2019gtm}.
We include elastic SD scattering only on hydrogen, the most abundant solar element, taking the contributions from other elements to be negligible.
For elastic SI scattering, we assume equal proton and neutron couplings, include all nuclear species in our abundance table, and use Helm form factors.

Thus, in this work we explore how inelastic SI, elastic SD, and elastic SI Higgsino scattering on solar nuclei affect DM cooling, the annihilation rate, and ultimately the IceCube constraints on Higgsino DM.
Given the theoretical uncertainty in the elastic SI cross section, we consider both the benchmark value introduced above and the case in which elastic SI scattering is neglected.


To model cooling through elastic scattering, we follow Ref.~\cite{Pospelov:2026ewn}.
We assume that Higgsinos oscillate along radial orbits while losing energy through elastic collisions.
The outer turning point of the orbit, $r_{\mathrm{max}}(t)$, therefore decreases with time according to
\begin{equation}
\label{eq:drmax}
    \frac{dr_{\mathrm{max}}}{dt}
    =\frac{1}{g(r_{\mathrm{max}})}
    \left\langle\frac{dE}{dt}\right\rangle\,,
\end{equation}
where $E=v^2/2+\Phi(r)$ is the orbital energy per unit Higgsino mass, $\Phi(r)$ is the gravitational potential per unit mass, and $\langle dE/dt\rangle$ is its change due to elastic scattering, averaged over one radial period.
The gravitational acceleration is $g(r)=GM(r)/r^2$, where $G$ is Newton's gravitational constant and $M(r)$ is the solar mass enclosed within radius $r$.
During this cooling stage, $\langle dE/dt\rangle<0$, so elastic scattering progressively reduces $r_{\mathrm{max}}(t)$, confining the captured Higgsinos to smaller orbits.

{\it  Thermalization:}
If captured Higgsinos have sufficient time to exchange energy with the solar plasma through elastic scattering, they can approach kinetic equilibrium and thermalize with the solar core, at which point net cooling ceases.
However, 
in the stationary-nucleus approximation of our cooling formalism, Eq. \eqref{eq:drmax} implies that Higgsinos can continue to lose energy and cool down to
arbitrarily small radii $r_{\mathrm{max}}(t)$.
To approximate the effect of thermalization, we therfore impose by hand that $r_{\mathrm{max}}(t)$ cannot become smaller than the characteristic thermal radius  adopted in Ref.~\cite{Pospelov:2026ewn},
\begin{equation}
    R_{\mathrm{therm}}=0.0037R_\odot\left(\frac{1~\mathrm{TeV}}{m_\chi}\right)^{1/2}\;,
\end{equation}
where $R_\odot$ is the solar radius.  

To determine whether thermalization occurs in our approximate treatment, we compare the cooling time, i.e. the time  required for the captured DM population to cool down to $R_{\mathrm{therm}}$, with the age of the Sun.  
Specifically, we define the cooling time $t_{\mathrm{cooling}}$ as the time required for the representative orbit to contract from 
$R_{\mathrm{stall}}$ to this floor:
\begin{equation}
    t_{\mathrm{cooling}}=
    \int_{R_{\mathrm{therm}}}^{R_{\mathrm{stall}}}
    \frac{g(r)}{|\langle dE/dt\rangle_r|}\,dr\;.
\end{equation}
The subscript $r$ denotes an orbit with outer turning point $r$; this time excludes the preceding inelastic-cooling stage.
Taking the Sun's age to be $t_\odot=4.5~\mathrm{Gyr}\simeq1.42\times10^{17}~\mathrm{s}$, we use $t_{\mathrm{cooling}}<t_\odot$ as an approximate thermalization criterion for the representative orbit.
If this criterion is satisfied, we set the population radius to $R=R_{\mathrm{therm}}$; otherwise, we take $R=r_{\mathrm{max}}(t_\odot)$.
In either case, replacing the population by a uniform sphere of radius $R$ is a separate approximation, not a consequence of thermalization or radial motion.

The limitations of this approximate treatment of cooling and thermalization must be considered when interpreting our results.
A particle captured at time $t_{\mathrm{cap}}$ has only $t_\odot-t_{\mathrm{cap}}$ available for cooling.
When $t_{\mathrm{cooling}}\ll t_\odot$, a broad range of capture times allows particles to reach the imposed thermal-radius floor, although the most recent captures remain unrelaxed.
This hierarchy reduces the importance of a calculation including finite capture ages.
The regime $t_{\mathrm{cooling}}\ll t_\odot$ was considered in Ref.~\cite{Nguyen:2026lui}, where for a thermal Higgsino with a mass of $1.1\ \mathrm{TeV}$, the authors obtained a cooling time of $\sim10^{16}\ \mathrm{s}$.
We also recover this cooling time for an $1.1\ \mathrm{TeV}$ Higgsino when both elastic SD and SI interactions contribute appreciably to cooling.
When $t_{\mathrm{cooling}}\approx t_\odot$, however, not all captured particles have sufficient time to reach the thermal-radius floor or, if that floor is not reached, even to contract to $r_{\mathrm{max}}(t_\odot)$.
Our treatment assigns all particles the radius obtained after a full solar age, irrespective of their capture times.
Within this radius prescription, that assumption can overestimate the central concentration and annihilation rate, potentially strengthening the inferred bounds.
The direction of the total error is not established, however, because the assumed uniform DM distribution and omitted nuclear thermal motion introduce additional uncertainties.
Finally, $t_{\mathrm{cooling}}\gg t_\odot$ does not by itself imply negligible elastic cooling: substantial contraction can occur before a slow final approach to the thermal-radius floor.
We therefore retain the calculated $r_{\mathrm{max}}(t_\odot)$ in this regime and use $R\approx R_{\mathrm{stall}}$ only when elastic interactions are completely absent or produce negligible contraction over the solar age.

\subsection{DM annihilation and capture--annihilation equilibrium}

As Higgsinos accumulate in the Sun, they can annihilate with one another.
We calculate the velocity-averaged Higgsino annihilation cross section $\langle\sigma v\rangle_{\mathrm{ann}}$ following the method in Ref.~\cite{Rinchiuso:2020skh} (Eqs.~6, 7, and 8), including the Sommerfeld enhancement.\footnote{For nonrelativistic Higgsinos heavier than the weak gauge bosons, 
the usual perturbative expansion for the annihilation cross section fails, 
and we must take into account the enhanced annihilation due to the Sommerfeld effect.}
To perform the velocity average, for nonthermalized populations, we assume that Higgsinos at radius $r<R$ have speed $v(r)=\sqrt{v_{\rm esc}^2(r)-v_{\rm esc}^2(R)}$ (obtained via energy conservation) and independent isotropic velocity directions, yielding the 
probability distribution for the relative velocity at a given r to be
 $p(v_{\rm rel}\mid r)=v_{\rm rel}/[2v^2(r)]$, where $v_{\rm rel}\in[0,2v(r)]$ is the relative velocity between two Higgsinos, and $R$ is the radius of the uniform DM distribution.
We then average the annihilation cross section over the uniform DM spherical population:
\begin{equation}
\langle\sigma v\rangle_{\mathrm{ann}}
=
\frac{3}{R^3}
\int_0^R r^2\,dr
\int_0^{2v(r)} dv_{\rm rel}\,
p(v_{\rm rel}\mid r)\,
(\sigma v_{\rm rel})\;.
\end{equation}
For populations treated as thermalized, we instead use a Maxwell--Boltzmann relative-speed distribution at the solar core temperature $k_BT_c=1.3~\mathrm{keV}$.
The Sommerfeld enhancement depends, in addition to the Higgsino mass, relative speed, and mass splitting $\delta$, on the mass difference between the charged Higgsinos and the lighter neutral Higgsino that constitutes the DM, $\Delta_+\equiv m_{\widetilde H^\pm}-m_{\widetilde H_1}$.
We adopt the benchmark $\Delta_+=350~\mathrm{MeV}$, also used in Ref.~\cite{Rinchiuso:2020skh}, which approximately corresponds to the radiative splitting in the heavy, nearly pure Higgsino limit and is consistent with the heavy-gaugino realization of Eq.~\ref{eq:higgsino_splitting} for $\delta$ of a few hundred keV.
Our method recovers the benchmark value $\langle\sigma v\rangle_{\mathrm{ann}}\simeq1.3\times10^{-26}~\mathrm{cm^3\; s^{-1}}$ at $m_\chi\simeq1.1~\mathrm{TeV}$ reported in Refs.~\cite{Beneke:2014hja,Dessert:2022evk} and used in Ref.~\cite{Pospelov:2026ewn}. 
We also find Sommerfeld resonances\footnote{These resonances arise when the electroweak potential supports a two-particle bound state close to the neutral-Higgsino pair threshold, strongly enhancing annihilation at low relative speeds~\cite{Mahbubani:2019shd}. 
Our perturbative treatment of annihilation does not include the finite annihilation widths needed to regulate sufficiently strong resonances.
Where the predicted rate approaches the partial-wave unitarity limit, a treatment that consistently includes these effects is required~\cite{Blum:2016nrz}.
The strongest resonance features should therefore be interpreted with this limitation in mind.} at Higgsino masses of approximately $6.5$, $25$, and $55~\mathrm{TeV}$, consistent with the resonance pattern shown in Refs.~\cite{Rinchiuso:2020skh,Mahbubani:2019shd}.

Equilibrium between capture and annihilation can be achieved when the capture rate in the Sun equals the rate of annihilation, and should not be confused with kinetic equilibrium.
Capture--annihilation equilibrium is reached if  $t_{\mathrm{eq}}\ll t_\odot$,
where $t_{\mathrm{eq}}$ is the characteristic timescale for establishing equilibrium between DM capture and annihilation,
\begin{equation}
    t_{\mathrm{eq}}=\sqrt{\frac{V_{\mathrm{eff}}}{C\langle\sigma v\rangle_{\mathrm{ann}}}}\;,
    \qquad V_{\mathrm{eff}}=\frac{4}{3}\pi R^3\; \,
    \label{t_eq}
\end{equation}
where $R$ is the radius of the uniform sphere within which we take the DM population to be uniformly distributed as described above.
If capture--annihilation equilibrium is achieved, the steady-state rate $\Gamma_{\mathrm{ann}}=C/2$ results, with a factor of 2 since each annihilation removes two particles. Below we will illustrate results with and without the assumption of capture--annihilation equilibrium. 

For constant capture rate $C$, negligible initial population, and negligible evaporation, the number of captured particles $N$ obeys 
\begin{equation}
\dot N=C-C_A N^2 \, , 
\end{equation}
with the constant coefficient $C_A=\langle\sigma v\rangle_{\mathrm{ann}}/V_{\mathrm{eff}}$~\cite{Nguyen:2026lui}.
The present-day DM annihilation rate is then
\begin{equation}
    \Gamma_{\mathrm{ann}}(t_\odot)=\frac{C}{2}\tanh^2\left(\frac{t_\odot}{t_{\mathrm{eq}}}\right)\;.
\end{equation}
As expected, for if  $t_{\mathrm{eq}}\ll t_\odot$, one finds $\Gamma_{\mathrm{ann}}=C/2$.

Let us clarify how cooling and thermalization influence the Higgsino annihilation rate in the Sun, which we later compare with the IceCube limits. 
When both SI and SD elastic scattering are neglected, cooling proceeds only through inelastic SI scattering and stalls at a radius $R=R_{\mathrm{stall}}$.
The resulting relatively extended DM population can have a large effective volume and a suppressed annihilation rate~\cite{Pospelov:2026ewn,Nguyen:2026lui}.
Further cooling through elastic scattering reduces the effective volume and increases the annihilation rate.
If cooling is effective enough, the capture--annihilation equilibrium condition can be satisfied, yielding the largest annihilation rate for a given capture rate. 
In the thermalized limit, which we approximate by setting $R=R_{\mathrm{therm}}$, we find that capture--annihilation equilibrium is reached throughout the Higgsino parameter region relevant to the IceCube constraints considered in this work.  The converse is not necessarily true: capture--annihilation equilibrium can be achieved with or without thermalization. 


\subsection{IceCube constraints on Higgsino DM}

We now compare the predicted Higgsino annihilation rates in the Sun with IceCube limits on the DM annihilation rate to constrain the mass splitting as a function of the Higgsino mass, considering both the thermal relic benchmark ($m_\chi\simeq1.1~\mathrm{TeV}$) and nonthermal scenarios at other masses.  
Following Refs.~\cite{Pospelov:2026ewn,Nguyen:2026lui}, we reconstruct the annihilation-rate limit $\Gamma_{\mathrm{lim}}$ from IceCube's published limits for the $W^+W^-$ channel~\cite{IceCube:2025fcu}.
We use this channel as a proxy for the Higgsino neutrino spectrum and exclude parameter points for which $\Gamma_{\mathrm{ann}}>\Gamma_{\mathrm{lim}}$.
Above $m_\chi=10~\mathrm{TeV}$, we hold $\Gamma_{\mathrm{lim}}$ fixed at its $10~\mathrm{TeV}$ value, following the benchmark extrapolation of Ref.~\cite{Langhoff:2026ujr}.

In Fig.~\ref{fig:higgsino_IceCube_limits}, we show four variants of the IceCube bounds on the Higgsino parameter space $(\delta, m_\chi$).
The strongest bounds, shown as gold dashed curves, are found when we impose capture--annihilation equilibrium for all Higgsino masses considered, so that $\Gamma_{\rm{ann}} = C/2$. 
Imposing the thermal-radius floor (plugging $R_{\rm therm}$ into Eq.~\ref{t_eq}) gives approximately the same bound within the Higgsino parameter space explored here.
With the assumption of capture--annihilation equilibrium, for the thermal Higgsino scenario ($m_\chi=1.1~\mathrm{TeV}$), we obtain $\delta\gtrsim566~\mathrm{keV}$, consistent with the $566~\mathrm{keV}$ bound at $1.08~\mathrm{TeV}$ in Ref.~\cite{Pospelov:2026ewn} and the thermal-Higgsino result in Ref.~\cite{Nguyen:2026lui}.
However, as discussed above, the capture--annihilation equilibrium condition should be checked separately for each Higgsino mass, since the resulting IceCube bounds are weaker when equilibrium is not reached.  

For the other three variants of IceCube bounds presented here, instead of imposing capture--annihilation equilibrium by hand, we check whether or not it has been reached under three different assumptions about the Higgsino--nucleon interactions that contribute to DM cooling. 
These determine how far the DM particles sink towards the center of the Sun, their final radii, and hence the annihilation rate.  

The next strongest bounds, shown by the green curves with circles, include cooling through all three Higgsino--nucleus interactions considered here: inelastic SI, elastic SI, and elastic SD scattering.
We fix the elastic SI cross section at $\sigma_{\mathrm{SI,elastic}}=4\times10^{-50}~\mathrm{cm}^2$, while the elastic SD cross section decreases as $m_\chi^{-2}$ in the large-mass limit.
Allowing for incomplete cooling, 
capture--annihilation equilibrium persists up to masses of approximately $10~\mathrm{TeV}$.
At larger masses, the weakening of the elastic SD interaction contributes to less efficient cooling, and the annihilation rate falls below $C/2$ which corresponds to capture--annihilation equilibrium.
At $m_\chi=10~\mathrm{TeV}$, we obtain $\delta\gtrsim510~\mathrm{keV}$, similar to the bound inferred in Ref.~\cite{Nguyen:2026lui}, where the authors assume a population thermalized with the solar core.
Sommerfeld resonances enhance the Higgsino annihilation cross section at specific masses. 
Around these masses, the IceCube bounds can become stronger but cannot exceed those obtained by imposing capture--annihilation equilibrium.
This strengthening is visible in the green curves with circles near the resonant masses of approximately $25$ and $55~\mathrm{TeV}$.

The elastic SI cross section is, however, subject to substantial theoretical uncertainties, and its contribution to cooling may be negligible.
We therefore consider a third case, shown by the purple solid curves, in which, in addition to inelastic SI scattering, SD scattering is the only elastic interaction contributing to cooling.
In this case, cooling is less efficient,  
and capture--annihilation equilibrium holds only up to masses of approximately $0.9~\mathrm{TeV}$.
Thus, for larger masses, elastic cooling only via the elastic SD interaction yields weaker IceCube limits on Higgsino DM than the case including also the elastic SI cooling channel.
For the thermal Higgsino benchmark, $m_\chi=1.1~\mathrm{TeV}$, we obtain $\delta\gtrsim558~\mathrm{keV}$.
Since the time required to cool to the thermal radius exceeds the solar age at the thermal Higgsino benchmark, this bound is sensitive to the approximations used to describe the captured population and its evolution.
In this case, elastic SD cooling becomes negligible over the solar age for Higgsino masses above approximately $5~\mathrm{TeV}$, and the IceCube bounds approximately coincide with the least restrictive bounds obtained with inelastic SI cooling alone which are shown in the magenta dot-dashed curve and discussed in the next paragraph.
Since all Sommerfeld resonances considered here occur above approximately $5~\mathrm{TeV}$, the IceCube bounds including both inelastic SI and elastic SD cooling behave approximately the same near these resonances as the bounds obtained with inelastic SI cooling alone, discussed in the next paragraph.

The least restrictive bounds, shown in Fig.~\ref{fig:higgsino_IceCube_limits} by the magenta dot-dashed lines, neglect elastic scattering altogether.
Without these loop-induced elastic interactions, cooling stalls at $R_{\mathrm{stall}}$.
Near the thermal-Higgsino exclusion boundary, the resulting extended population reaches neither kinetic equilibrium with the solar core nor capture--annihilation equilibrium in our model.
For the thermal Higgsino benchmark, the resulting bound is $\delta\gtrsim506~\mathrm{keV}$, consistent with the result at $1.08~\mathrm{TeV}$ in Ref.~\cite{Pospelov:2026ewn}.
For masses below approximately $5~\mathrm{TeV}$, this case is extreme because it neglects the loop-induced elastic interactions entirely. 
However, for masses above approximately $5~\mathrm{TeV}$, as discussed in the previous paragraph, the magenta dot-dashed and purple solid curves are essentially the same, so that the bounds neglecting elastic scattering  approximately coincide with those obtained by considering both inelastic SI and elastic SD cooling, which is a more realistic scenario. 
At $m_\chi=10~\mathrm{TeV}$, for instance, we obtain $\delta\gtrsim390~\mathrm{keV}$, substantially weaker than the approximately $510~\mathrm{keV}$ bound inferred from Ref.~\cite{Nguyen:2026lui}, which assumes a population thermalized with the solar core.
The Sommerfeld resonance near $m_\chi\simeq6.5~\mathrm{TeV}$ also produces a visible strengthening of these bounds.

Our estimates of the IceCube limits for Higgsino masses between $0.1$ and $200~\mathrm{TeV}$ illustrate their strong dependence on the assumed elastic interactions and the evolution of the captured population.
Our cooling-time estimates should nevertheless be interpreted with caution, as our calculation treats solar nuclei as stationary.
In particular, protons in the solar core have appreciable thermal velocities, which can increase the elastic SD collision rate relative to the stationary-target estimate when Higgsino speeds are small, without changing the underlying cross section.
Scattering from a moving nucleus can transfer energy either to or from a Higgsino, and a consistent treatment of kinetic relaxation must include both processes.
The stationary-nuclei approximation, together with the radial-orbit, uniform-density, and fixed-effective-volume assumptions, limits the quantitative reliability of our cooling-dependent bounds, especially when the cooling time is comparable to the solar age.

To summarize, we found four variants of IceCube bounds on Higgsino parameter space.
The most constraining gold-dashed curve makes the unrealistic assumption of s capture--annihilation equilibrium for all Higgsino masses, whereas the other three variants actually compute the annihilation rate under three different assumptions about the Higgsino--nucleon scattering, which determines the final radius to which the Higgsinos cool and thus the final annihilation rate.  
The three variants of bounds we find are: the green curve when we include all three types of interactions (SI inelastic, SD elastic, and SI elastic), the purple solid curve when we include inelastic and SD elastic, and the magenta dot-dashed curve when we include only the inelastic scattering.  We note that the purple solid and magenta dot-dashed curves are essentially identical for Higgsino masses above 5 TeV. We believe the gold-dashed curve above 10 TeV and the 
 magenta dot-dashed curve below 5 TeV to be too extreme to be realistic: the gold-dashed curve assumes capture--annihilation equilibrium for all masses and the magenta dot-dashed curve does not take into account any elastic scattering at all. The green and purple solid curves outline reasonable  bounds on Higgsino parameter space. 


\section{ Results}

In this section, we compare the Higgsino parameter space compatible with the LZ event of interest~\cite{LZ:2026axp} with the IceCube constraints derived from Higgsino capture and annihilation in the Sun. The regions compatible with these experiments are shown in Fig.~\ref{fig:higgsino_IceCube_limits} in the $(\delta,m_\chi)$ plane.

Building on our earlier paper~\cite{Freese:2026sga}, we extend our analysis to Higgsino masses above the thermal relic benchmark ($m_\chi\simeq1.1~\mathrm{TeV}$).
Using the approximate likelihood described above, we perform a likelihood-ratio analysis to determine the 90\% CL regions in the $(\delta,m_\chi)$ plane, both with and without the high-energy LZ sideband.
In Fig.~\ref{fig:higgsino_IceCube_limits}, the orange regions show the results obtained without the sideband, while the blue regions include the constraint from the absence of events in this energy range.
The upper panel assumes the Standard Halo Model (SHM), while the lower panel includes the enhancement of the high-speed tail due to the LMC.
The IceCube bounds at 90\% CL discussed in the previous section are also shown for comparison.

We first consider the (unrealistic for masses above $10\ \mathrm{TeV}$) limiting  assumption of capture--annihilation equilibrium for all Higgsino masses, shown in the gold-dashed curve, which gives the strongest IceCube bounds for a given capture rate.  
Even under this assumption, regions compatible with both LZ and IceCube remain, whether or not the sideband is included.
For the SHM, the overlap lies at $m_\chi\gtrsim80~\mathrm{TeV}$, with $340~\mathrm{keV}\lesssim\delta\lesssim390~\mathrm{keV}$.
For the SHM + LMC distribution, it spans approximately $13~\mathrm{TeV}\lesssim m_\chi\lesssim120~\mathrm{TeV}$, with $465~\mathrm{keV}\lesssim\delta\lesssim520~\mathrm{keV}$.
Mass and splitting remain correlated within these regions.
Capture--annihilation equilibrium over the full Higgsino mass range is considered in Ref.~\cite{Langhoff:2026ujr}, which identifies regions compatible with both LZ and IceCube at Higgsino masses of order $100$--$1000~\mathrm{TeV}$ within its 68\% confidence contours.
This mass range does not by itself indicate a discrepancy with our results, since we present 90\% confidence regions.
Indeed, the broader $2\sigma$ contours in Ref.~\cite{Langhoff:2026ujr}, corresponding to 95\% CL, allow lower masses and show qualitative agreement with our 90\% confidence regions when capture--annihilation equilibrium is imposed.


We next compare the LZ contours with the IceCube limits obtained by accounting for different cooling channels without imposing capture--annihilation equilibrium.
As discussed in Sec.~\ref{sec:icecube}, the assumed interactions contributing to cooling can significantly affect the IceCube limits. 
The three interactions considered are inelastic SI scattering, which governs Higgsino--nucleus scattering in xenon-based detectors, and elastic SD and SI scattering, which are subdominant in direct detection.

The green dot-dashed curve shows our results when we include all three interactions in the cooling.
For the SHM, compatibility with LZ and IceCube requires approximately $m_\chi\gtrsim30~\mathrm{TeV}$, with mass splittings roughly between $350$ and $400~\mathrm{keV}$, regardless of whether the high-energy sideband is included.
For SHM + LMC, the compatible region differs only little from that obtained by imposing capture--annihilation equilibrium which is discussed in the previous paragraph.

We then explore a scenario in which the elastic SI interaction is negligible, as allowed by theoretical uncertainties~\cite{Chen:2019gtm}.
Here, elastic SD scattering plays a central role in cooling.
Its cross section scales as $\sigma_{\mathrm{SD,elastic}}\propto m_\chi^{-2}$ in the heavy-mass limit~\cite{Hisano:2011cs}\footnote{This dependence on the DM mass was missed in previous studies of IceCube bounds.} and therefore becomes considerably weaker above the thermal Higgsino mass, increasing the cooling time and weakening the IceCube limits.
For the SHM + LMC velocity distribution, Higgsinos with masses as low as approximately $2~\mathrm{TeV}$ and $\delta\sim520~\mathrm{keV}$ remain compatible with both experiments (LZ and IceCube) at 90\% CL if the high-energy sideband is omitted.
For the SHM, the overlap occurs at $m_\chi\gtrsim10~\mathrm{TeV}$, with mass splittings roughly between $370$ and $400~\mathrm{keV}$, whether or not the sideband constraint is included.

Finally, we consider cooling through inelastic SI scattering alone, neglecting both elastic channels.
This scenario is not realistic for Higgsino masses below $5\ \mathrm{TeV}$ where the elastic scattering cannot be neglected.
In this limiting case, the thermal Higgsino is excluded for the SHM but remains marginally compatible with the IceCube limits and the LZ event at 90\% CL for SHM + LMC and $\delta\sim508~\mathrm{keV}$, provided the high-energy sideband is omitted.  Hence this is the only case in which the thermal Higgsino remains a viable explanation of the LZ data.
In summary, regions compatible with both LZ and IceCube remain for all cooling assumptions considered here, whether or not the high-energy LZ sideband is included.
Depending on the halo distribution and cooling interactions, these regions occur at Higgsino masses from $\sim2\ \mathrm{TeV}$ to tens of TeV or higher.
Obtaining the observed DM abundance at these masses requires an abundance history beyond standard radiation-dominated freeze-out~\cite{Langhoff:2026ujr}.

Although the 1.1 TeV thermal Higgsino is only marginally viable in the extreme case where all elastic scattering cooling mechansims are neglected,
a Higgsino mass as low as $\sim2\ \mathrm{TeV}$ remains a good candidate. It provides an explanation of the LZ event and remains consistent with IceCube bounds when we make reasonable assumptions about cooling, when LMC  enhancement is included (but ignoring the empty LZ sideband).

\section{Discussion}
In this paper, we have explored thermal and nonthermal Higgsino dark-matter interpretations of the $248\pm23\;(\mathrm{stat})\pm23\;(\mathrm{syst})~\mathrm{keV}$ nuclear-recoil event of interest observed by LZ~\cite{LZ:2026axp}, expanding upon our earlier work~\cite{Freese:2026sga}.
The event lies in a region of low expected background, and none of the rare background processes or detector effects investigated by the LZ Collaboration was identified as a likely explanation.
In this work, we have considered the enhancement of the local DM high-speed tail due to the Large Magellanic Cloud (LMC), the constraint from the absence of events in the high-energy LZ sideband, and DM-annihilation constraints from IceCube under different assumptions about Higgsino cooling and annihilation in the Sun.
Our main results are summarized in Fig.~\ref{fig:higgsino_IceCube_limits}, where regions in the Higgsino parameter space compatible with both the LZ event and IceCube bounds are presented.

We find that the thermal Higgsino with $m_\chi\simeq1.1~\mathrm{TeV}$ is incompatible with the combination of LZ and IceCube if we assume the Standard Halo Model (SHM) only.
For the SHM + LMC distribution which accounts for the enhanced high-speed tail of the DM velocity distribution, thermal Higgsino remains marginally compatible near $\delta\simeq508~\mathrm{keV}$ only when both SI and SD elastic cooling channels are neglected and the high-energy sideband is omitted.
Since the loop-induced elastic SD interaction contributes considerably to cooling at $m_\chi\simeq1.1~\mathrm{TeV}$, this possibility represents a limiting case rather than a robust thermal-Higgsino interpretation.

Heavier Higgsinos allow compatibility with both datasets at 90\% confidence level without the above restrictive assumptions needed in the thermal Higgsino interpretation.
Depending on the halo distribution and cooling channels, the surviving regions extend from Higgsino masses $\sim2\ \mathrm{TeV}$ to above several hundred TeV.  Although the  parameter range we studied only went up to Higgsino masses of 200 TeV, it is clear from our results that higher mass Higgsinos will be compatible with LZ and IceCube bounds as well.
The relevant splittings are approximately $350$--$400~\mathrm{keV}$ for the SHM and $460$--$525~\mathrm{keV}$ for SHM + LMC, with the mass and splitting correlated within each region.
Compatible regions remain even in the most extreme case considered, when the high-energy sideband is included and capture--annihilation equilibrium (which yields the strongest IceCube limits at a fixed capture rate but is unrealistic for $m_{\chi} > 10$ TeV)  is imposed.
Obtaining the observed DM abundance at these masses requires an abundance history beyond standard radiation-dominated freeze-out~\cite{Langhoff:2026ujr}.

The DM-annihilation constraints depend on whether the captured Higgsinos cool efficiently enough to approach capture--annihilation equilibrium.
The decrease of the elastic SD cross section as $m_\chi^{-2}$ in the heavy-mass limit~\cite{Hisano:2011cs} can substantially reduce cooling at large masses.
Consequently, imposing equilibrium throughout the parameter space can overestimate the annihilation rate.
We also show that Sommerfeld enhancement can restore stronger bounds near resonant masses, but, within our population-evolution model, these bounds cannot exceed those obtained when we have capture-annihilation equilibrium for a fixed capture rate.

Several uncertainties affect the comparison with LZ.
Inelastic scattering at the splittings considered here probes the high-speed tail of the local DM distribution, making the inferred parameter regions sensitive to its shape and normalization.  We took into account the effects of the LMC on the high velocity tail by using a particular model (``halo-13") in the work of \cite{Smith-Orlik:2023kyl}.
As discussed in Section \ref{sec:LMC}, the SHM  + LMC distribution remains uncertain; further, other massive bodies in the MW may have effects not yet taken into account.

Another question is whether or not to impose the constraint of an empty high energy recoil LZ sideband on our model.
The sideband constraint  depends on the assumed signal acceptance, which has not been published for this region.
Our approximate likelihood therefore cannot replace a dedicated analysis by the LZ Collaboration. 
The analysis in this paper is performed with and without taking into account the empty sideband.

The solar capture calculation has additional uncertainties associated with the abundance and radial distribution of heavy elements.
For species without individual radial profiles in our adopted solar model, we assume constant mass fractions (an assumption also adopted in Ref.~\cite{Pospelov:2026ewn}), so that their number densities follow the total solar density~\cite{Pospelov:2026ewn}.
Although their normalizations are based on the abundances in Ref.~\cite{Asplund_2021}, these inputs do not determine their interior profiles.
This approximation matters because heavy nuclei provide the targets needed to overcome the inelastic threshold.
Nuclear form-factor uncertainties can affect both Higgsino capture and cooling, with their impact on elastic scattering illustrated in Ref.~\cite{Chen:2019gtm}.
These effects are distinct from the scalar--spin-2 cancellation in the loop-induced elastic SI amplitude~\cite{Chen:2019gtm}, which significantly affects Higgsino cooling in the Sun and consequently the IceCube bound on Higgsino DM.
The combined impact of these uncertainties on the inferred bounds has not been quantified.

The evolution of the captured population is another important approximation.
We describe cooling through a characteristic orbital radius and approximate the final population as a uniform sphere.
Our annihilation calculation uses the resulting final spatial and velocity distributions rather than evolving cooling and annihilation together for particles captured at different times.
When cooling is slow, assigning the final distribution to the entire population can overestimate its central concentration and thereby lead to strengthen the inferred
IceCube bounds.
A calculation that follows the spatial and velocity distributions throughout capture, cooling, and annihilation would provide a more reliable treatment of these regions.

Finally, our conversion of the predicted annihilation rate into an IceCube constraint uses the published $W^+W^-$ channel as an approximation to the Higgsino neutrino signal.
A dedicated calculation would account for the Higgsino annihilation branching fractions, neutrino propagation through the Sun, and the detector response.
The IceCube limits on the $W^+W^-$ channel used here extend to $10~\mathrm{TeV}$~\cite{IceCube:2025fcu}; above this mass, we hold the inferred annihilation-rate limit fixed at its $10~\mathrm{TeV}$ value.
The resulting high-mass bounds and compatibility regions are therefore benchmark extrapolations.
Establishing their quantitative extent requires a neutrino-signal analysis covering those masses.

Complementary observations can further test the Higgsino DM interpretation.
Super-Kamiokande data also constrain neutrinos from solar Higgsino annihilation, although the limits reported in Ref.~\cite{Bose:2026ndd} are weaker than those from IceCube.
Paleodetectors are a proposed experimental technique for direct detection of DM via the readout of DM-induced nuclear recoil tracks in natural minerals \cite{Theodosopoulos:2026ehn,Theodosopoulos:2026eym,Theodosopoulos:2026kgy,Elykov:2026ojl}. 
Heavy-element paleodetectors offer an independent probe of inelastic Higgsino scattering through recoil tracks in ancient minerals~\cite{Graham:2026ivn}.
Together with additional high-energy recoil data and improved solar-neutrino calculations, such searches can test the parameter regions identified here.

\acknowledgments
We thank Dan Akerib, Rick Gaitskell, Chris Kelso, Pat Stengel, Monica Valluri and Sunny Vagnozzi for helpful discussions. KF is grateful for support from the Jeff \& Gail Kodosky Endowed Chair in Physics at the University of Texas. KF acknowledges support from the Swedish Research Council (Contract No. 638-2013-8993).  KF and DT acknowledge support by the U.S. Department of Energy, Office of Science, Office of High Energy Physics program under Award Number DE-SC-0022021.

\bibliography{refs}


\end{document}